\documentclass[aps,prb,twocolumn,amsmath,amssymb,superscriptaddress,floatfix,citeautoscript*]{revtex4-1}

\usepackage{graphicx, multirow, xcolor, bm, ulem}
\usepackage{amsfonts,amssymb,amsmath}
\usepackage{graphicx,dcolumn,bm,color,braket,slashed}
\usepackage{times, comment, mathtools, extarrows}
\usepackage{appendix}

\begin{document}

\title{
Topological and trivial domain wall states in engineered atomic chains
}

\author{Seung-Gyo Jeong}
\affiliation{Department of Physics, Pohang University of Science and Technology (POSTECH), Pohang 37673, Korea}

\author{Tae-Hwan Kim}
\email{taehwan@postech.ac.kr}
\affiliation{Department of Physics, Pohang University of Science and Technology (POSTECH), Pohang 37673, Korea}

\maketitle

In a recent article~\cite{huda2020}, Huda \textit{et al}.\ demonstrated tuneable topological domain wall states in the $c$(2$\times$2) chlorinated Cu(100).
Their system~\cite{drost2017} allows to experimentally tune the domain wall states using atom manipulation by the tip of a scanning tunneling microscope (STM).
They have realized topological domain wall states of two prototypical 1D models such as trimer~\cite{alvarez2019} and coupled dimer chains~\cite{kim2012,cheon2015,kim2017,han2020,oh2021}.
However, they did not distinguish trivial domain wall states~\cite{blanco2016} from topological ones in their models.
As a result, all states of a specific domain wall are not topological but trivial.
Here, we show why the specific domain wall states are trivial and how to make them topological.
This topological consideration would provide more clear insight on future studies on topological domain wall states in artificial atomic chains.

First, we point out a limitation of the discrete atomic lattice: artificially designed ground states do not respect the symmetry of target models. 
For example, in the Su–Schrieffer–Heeger (SSH) model, 
due to their $Z_2$ symmetry~\cite{han2020}, one ground state should be transformed to the other and then return to the initial ground state via two successive identical SSH domain walls ($A \rightarrow B \rightarrow A$) (Supplementary Fig.~1a).
However, the same transformation cannot be realized for  the chlorinated Cu(100) (Supplementary Fig.~1b).
If we use the same domain wall structure twice, then we end up with a different ground state ($A'$) from the initial one ($A$).
To correct this discrepancy, we need to introduce a somewhat different domain wall configuration indicated by the blue square in Supplementary Fig.~1c.
Thus, this corrected atomic configuration does not fully respect the SSH symmetry because two domain walls are not identical like the SSH model.

In a similar way, we consider the coupled dimer chain consisting of two coupled 1D dimerized chains.
Because two opposite chiral domain walls (red and blue dots in Supplementary Fig.~1d) on the same chain can annihilate each other, one ground state can return to itself ($AA \rightarrow BA \rightarrow AA$).
However, due to the limitation of the discrete atomic lattice, one ground state cannot come back to the initial ground state ($AA \rightarrow BA \rightarrow A''A''$, Supplementary Fig.~1e) by using opposite chiral domain walls proposed by Huda \textit{et al}.
To rectify this contradiction, we suggest another domain wall configuration that includes an additional atom (blue square in Supplementary Fig.~1f).

\begin{figure}[h]
\centering \includegraphics[width=86mm]{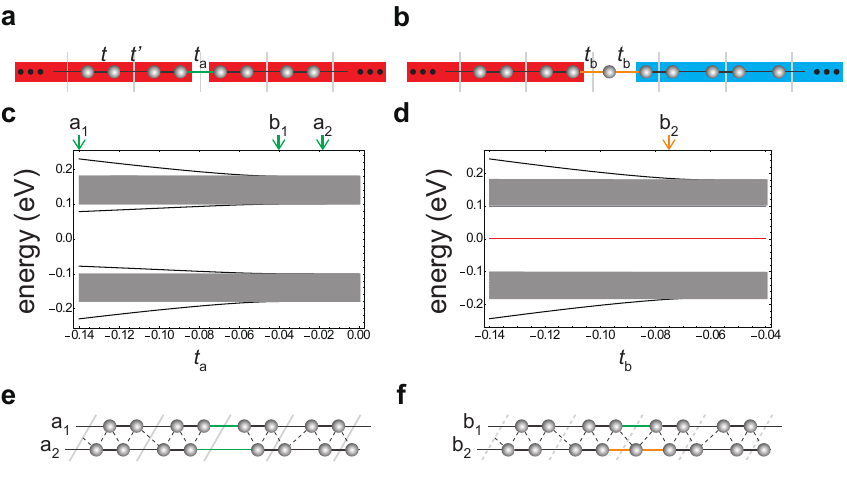}
\caption{\label{fig:1Dcentervariable}
\textbf{Topological and trivial domain walls.}
\textbf{a,b} Schematics of topologically distinct domain walls in dimerized SSH chains. 
$t_\text{a}$ (green) and $t_\text{b}$ (orange) represent variable hopping parameters at both domain walls 
while $t$ and $t'$ indicate intradimer and interdimer hopping parameters, respectively. 
Dimers with $t>t'$ ($t<t'$) are shaded in red (blue).
\textbf{c,d} Calculated energy diagrams of \textbf{a} and \textbf{b} as functions of $t_\text{a}$ and $t_\text{b}$.
Gray shaded regions indicate bulk states 
while red and black lines denote topological and trivial domain wall states, respectively.
\textbf{e,f} Schematics of topologically distinct domain walls in coupled dimer chains.
Both upper ($a_1$) and lower ($a_2$) dimer chains in \textbf{e} show topologically equivalent domain walls as \textbf{a}.
On the other hand, the the upper ($b_1$) and lower ($b_2$) dimer chains in \textbf{f} have topologically different domain walls, \textbf{a} and \textbf{b}, respectively. 
Corresponding energy diagrams of each dimer chain in \textbf{e} and \textbf{f} are indicated in \textbf{c} and \textbf{d}.
}
\end{figure}

Next, we discuss the topology of domain wall states in 1D chain systems.
In the SSH model, the topological invariants or Zak phases~\cite{zak1989} of dimer chains depend on the ratio between intradimer ($t$) and interdimer ($t'$) hopping parameters.
The Zak phase, $\theta_\text{Zak}$, can be obtained through the Bloch wave functions $|u_k \rangle$: $\theta_\text{Zak}=i  \int_{-\pi / a_0}^{\pi / a_0} \langle u_k |\partial_k u_k \rangle \,dk $, where $a_0$ is the lattice period of the dimer chain.
As shown in Fig.~\ref{fig:1Dcentervariable}a,b, a ground state shows $\theta_\text{Zak}=0$ when $|t/t'|>1$ (red regions) while $\theta_\text{Zak}=\pi$ when $|t/t'|<1$ (a blue region).
Topological domain wall states emerge at the interface between topologically distinct Zak phases ($\theta_\text{Zak}=0 \leftrightarrow \pi$, Fig.~\ref{fig:1Dcentervariable}b,d) 
while we do not expect any topological edge states between the topologically same Zak phases ($\theta_\text{Zak}=0 \leftrightarrow 0$, Fig.~\ref{fig:1Dcentervariable}a,c).

Such topological domain wall states depend only on topology of ground states but not the strength of hopping parameters~\cite{blanco2016}. 
In other words, adiabatic deformation of Hamiltonians does not change the topology of domain wall states. 
Therefore, all Hamiltonians are said to be adiabatically equivalent or adiabatically connected when we continuously tune the hopping parameters $t_a$ and $t_b$ at domain walls by maintaining the system's symmetry (Fig.~\ref{fig:1Dcentervariable}a,b). 
As shown in Fig.~\ref{fig:1Dcentervariable}c,d, both domain wall states do maintain their topology even though trivial edge states appear at higher hopping parameters at the domain walls ($|t_{a,b}| > min(|t|,|t'|)$).

In this sense, the domain wall $AA \rightarrow AB$ (Fig.~\ref{fig:1Dcentervariable}e) proposed by Huda \textit{et al}.\ actually does not have any topological edge modes because the upper and lower chains exhibit the same trivial topology such as Fig.~\ref{fig:1Dcentervariable}a.
In contrast, the domain wall $AA \rightarrow AB$ (Fig.~\ref{fig:1Dcentervariable}f) suggested by us indeed do have a topological edge mode at the lower chain.

\begin{figure}[ht]
\centering \includegraphics[width=86mm]{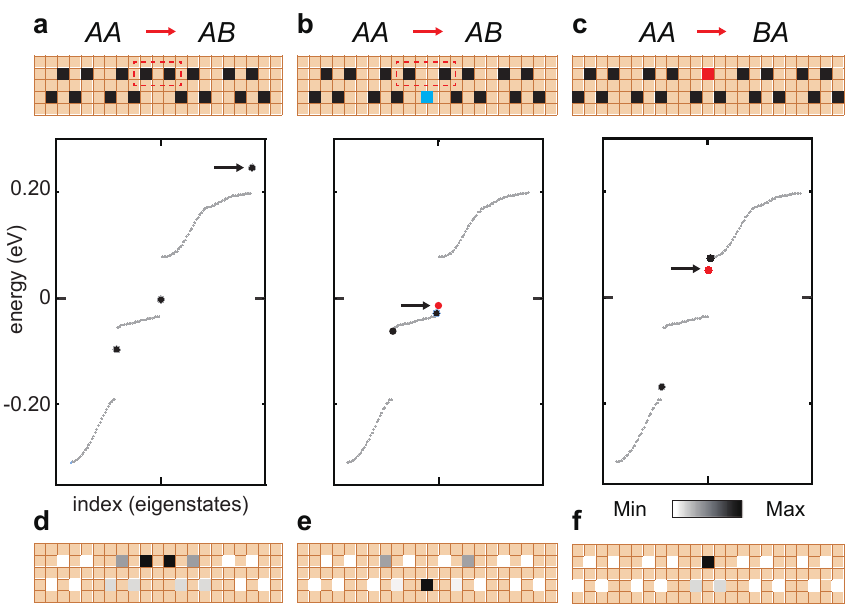} 
\caption{\label{fig:models}
\textbf{Topological and trivial domain wall states in coupled dimer chains.}
\textbf{a--c} Atomic configurations and corresponding energy spectra of domain walls: $AA \rightarrow AB$ (\textbf{a},\textbf{b})  and $AA \rightarrow BA$ (\textbf{c}).
The atomic configurations in \textbf{a} and \textbf{b} have the same domain walls as Fig.~1e and 1f, respectively.
Red dashed rectangles indicate the difference on the upper chains between \textbf{a} and \textbf{b}. 
Red, black, and gray dots in the energy spectra represent topological, trivial domain wall states, and bulk states, respectively. 
\textbf{d--f} Simulated local density of states (LDOS) maps of \textbf{a}--\textbf{c} at the energy levels indicated by black arrows in \textbf{a}--\textbf{c}. 
Black (white) denotes the highest (lowest) LDOS.
}
\end{figure}

Based on our topological consideration, we perform the tight-binding calculations as Huda \textit{et al}.\ did~\cite{huda2020} to compare two different domain walls.
As shown in Fig.~\ref{fig:models}a,b, two domain wall configurations exhibit distinct energy spectra.
Most notably, all domain wall states (black dots in Fig.~2a) are trivial in the domain wall Huda \textit{et al}.\ proposed while a topological domain wall state (red dot in Fig.~2b) emerges in the upper band gap in our domain wall configuration.
We confirm that such domain wall states are adiabatically connected to ones without interchain coupling as well as some trivial domain states appear under the strong interchain coupling as described in Supplementary Fig.~2.


Furthermore, we check the spatial localization of these domain wall states by plotting the simulated local density of states (LDOS) maps as shown in Fig.~\ref{fig:models}d--f.
Whereas the trivial domain wall state is localized at the upper chain where there is no phase shift ($A \rightarrow A$, Fig.~\ref{fig:models}d), the topological state mainly exists at the domain wall site on the lower chain where there is the distinct topology shift ($A \rightarrow B$, Fig.~\ref{fig:models}e).
Such topological properties are also observed in another chiral domain wall $AA \rightarrow BA$ reported by Huda \textit{et al}.\ (Fig.~\ref{fig:models}c,f).
Note that Fig.~\ref{fig:models}b and c now exhibit the topologically opposite chirality, which has been reported in other systems~\cite{cheon2015,kim2017,oh2021}.
On the other hand, Fig.~\ref{fig:models}a and c are not topologically comparable because one is trivial and the other is topological.

We showed why the original domain wall configuration is not topological and the new configuration has the topological domain wall state.
However, the new configuration is impossible to realize on the chlorinated Cu(100).
Because the domain wall should be located in between possible chlorine sites.
Instead, we propose an alternative configuration by maintaining the same topology (Supplementary Fig.~3).
Thus, we can experimentally access both topological and trivial domain states because they exist below the conduction band of the chlorine layer~\cite{huda2020}.

Although the authors have demonstrated various tuneable topological domain wall states using atom manipulation,
they have not properly considered topology.
As a result, one of the domain wall configurations proposed by Huda \textit{et al}.\ does not have any topological edge mode and is inconsistent with their other topological domain walls.
We suggested the alternative domain wall configuration, which recovers a topological domain wall state regardless of the interchain coupling.
Such careful topological considerations will provide further insight on topological domain wall states in any artificial atomic chains.

\begin{acknowledgments}
We thank Ha-Eum Kim for advising the figure data and Sangmo Cheon for useful discussions. This work was supported by the National Research Foundation of Korea (NRF) funded by the Ministry of Science and ICT, South Korea (Grants No. NRF-2021R1F1A1063263, NRF-2018R1A5A6075964, and 2016K1A4A4A01922028).
\end{acknowledgments}

\begin{description}
\item[Author contributions]
S.-G.J. and T.-H.K. conceived the idea, carried out the analyses, and wrote the article.

\item[Competing interests]
The authors declare no competing interests.

\item[Additional information]
\textbf{Correspondence and requests for materials} should be addressed to T.-H.K.

\end{description}

\bibliography{Reference.bib}
\end{document}


\title{Supplementary Information:\\Topological and trivial domain wall states in engineered atomic chains
}

\author{Seung-Gyo Jeong}
\affiliation{Department of Physics, Pohang University of Science and Technology (POSTECH), Pohang 37673, Korea}

\author{Tae-Hwan Kim}
\email{taehwan@postech.ac.kr}
\affiliation{Department of Physics, Pohang University of Science and Technology (POSTECH), Pohang 37673, Korea}

\maketitle

\setcounter{page}{1}

\begin{figure}[ht]
\centering \includegraphics[width=0.8\linewidth]{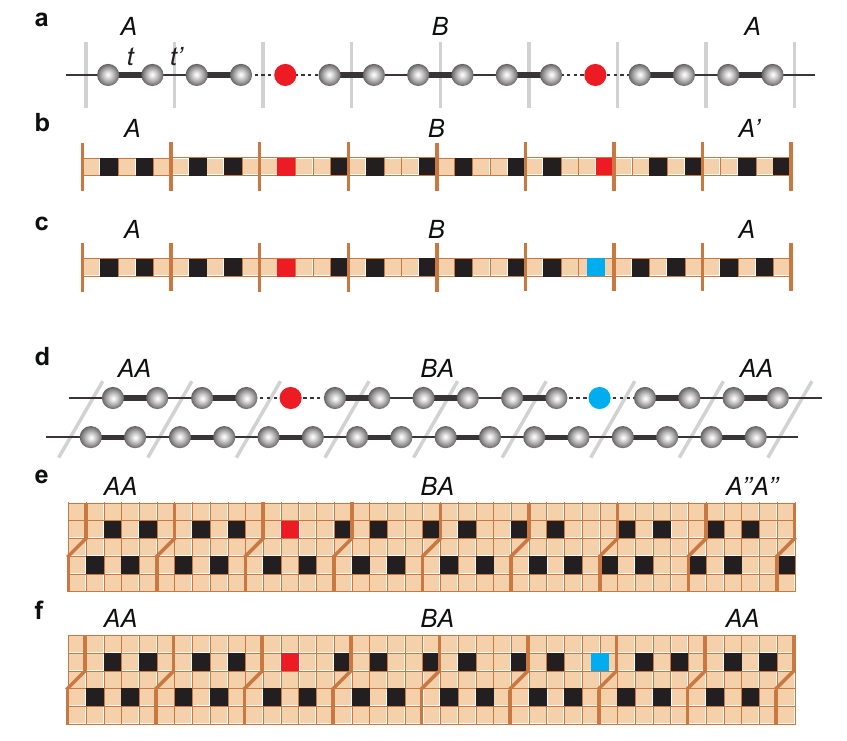} 
\caption{
\textbf{Atomic schematics of 1D domain walls.}
\textbf{a} Schematic model of an SSH chain having two identical domain walls.
Ground states can have either $A$ ($|t/t'|>1$) or $B$ ($|t/t'|<1$) phases.
Red atoms represent domain walls connecting one ground state to the other. 
\textbf{b,c} Atomic configurations on the chlorinated Cu(100) of \textbf{a}. 
Black and red/blue squares (chlorine vacancies) denote bulk and domain walls, respectively.
Two red domain walls connect $A$ to $A'$ in \textbf{b}
while red and blue domain walls do $A$ to $A$ in \textbf{c}.
\textbf{d} Schematic model of coupled dimer chains having two opposite chiral domain walls. 
Two domain walls (indicated by red and blue atoms) topologically annihilate each other.
\textbf{e,f} Atomic configurations on the chlorinated Cu(100) of \textbf{d}.
In \textbf{e}, an initial ground state $AA$ is transformed into a somewhat different ground state $A''A''$ by using two domain walls Huda \textit{et al}.\ proposed.
In contrast to \textbf{e}, a ground state returns to itself via two chiral domain walls in \textbf{f}.
Note that the blue domain wall cannot be experimentally realized on the chlorinated Cu(100) (see Fig.~S3).
}
\label{fig:Group}
\end{figure}

\newpage

\begin{figure}[h]
\centering\includegraphics[width=0.7\linewidth]{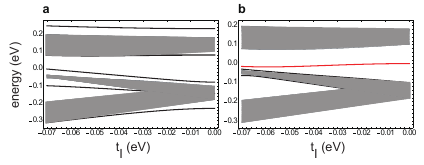}
\caption{\label{fig.cvariable}
\textbf{Adiabatic evolution of topological and trivial domain wall states.}
\textbf{a,b} Energy diagrams as a function of the interchain coupling ($t_\text{I}$) of Fig.~2\textbf{a,b}.
Red and black lines indicate topological and trivial domain wall states, respectively, while gray shaded regions denote bulk states.
Such adiabatic evolution maintains topology of domain wall states regardless of the interchain coupling strength.
The left and right ends represent Fig.~2 in the main text and ones without the interchain coupling, respectively.
}
\end{figure}
\begin{figure}[hb]
\centering \includegraphics[width=0.7\linewidth]{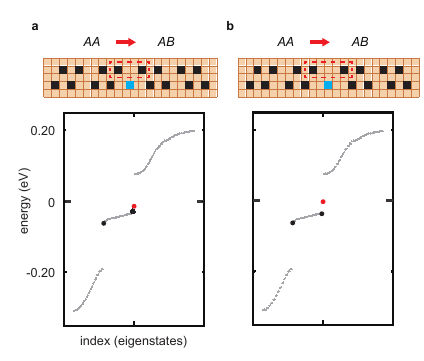}
\caption{\label{fig.Exp}
\textbf{Experimentally feasible AA--AB domain wall.}
\textbf{a} Proposed AA $\rightarrow$ AB domain wall in Fig.~2b of the main text.
\textbf{b} Alternative AA $\rightarrow$ AB domain wall for experimental realization by adding more spaces between two ground states. 
Red dashed rectangles highlight the difference of the upper chains between \textbf{a} and \textbf{b}.
Tight-binding calculations give similar energy spectra and confirm these two configurations are adiabatically connected (or topologically the same).
Furthermore, the topological domain wall state exists in the band gap, which is experimentally accessible because it is much lower than the conduction band of the chlorine layer.
}
\end{figure}

\bibliography{Reference.bib}